\documentclass[aps,twocolumn,pre]{revtex4}
\usepackage{bm}
\usepackage{braket}
\usepackage{amsfonts,amssymb}
\usepackage{epsfig, amsmath}
\usepackage{bm}
\usepackage{soul}
\usepackage{amsmath}
\usepackage{amssymb}
\usepackage{bm}
\usepackage{verbatim}
\usepackage{graphicx}
\usepackage{subfigure}
\usepackage{psfrag}
\usepackage{epsfig}

\begin{document}

\title{Ballistic thermophoresis of adsorbates on free-standing graphene}

\author{Emanuele Panizon$^{a}$, Roberto Guerra$^{a,b}$ and Erio Tosatti$^{a,c,d,1}$}
\affiliation{$^a$International School for Advanced Studies (SISSA), Via Bonomea 265, 34136 Trieste, Italy\\
$^b$Dipartimento di Fisica, Universit\`a degli Studi di Milano, Via Celoria 16, 20133 Milano, Italy\\
$^c$The Abdus Salam International Centre for Theoretical Physics (ICTP), Strada Costiera 11, 34151 Trieste, Italy\\
$^d$CNR-IOM Democritos National Laboratory, Via Bonomea 265, 34136 Trieste, Italy}

\begin{abstract} 
The textbook thermophoretic force which acts on a body in a fluid is proportional to the local temperature gradient. The same is expected to hold for the macroscopic drift behavior of a diffusive cluster or molecule physisorbed  on a solid surface.
The question we explore here is whether that is still valid on a 2D membrane such as graphene at short sheet length.
By means of a non-equilibrium molecular dynamics study of a test system -- a gold nanocluster adsorbed on free-standing graphene clamped between two temperatures $\Delta T$ apart -- we find a phoretic force which for submicron sheet lengths is parallel to, but basically independent of, the local gradient magnitude. This identifies a thermophoretic regime that is ballistic rather than diffusive, persisting up to and beyond a hundred nanometer sheet length. Analysis shows that the phoretic force is due to the flexural phonons, whose flow is known to be ballistic and distance-independent up to relatively long mean-free paths.
Yet, ordinary harmonic phonons should only carry crystal momentum and, while impinging on the cluster, should not be able to impress real momentum. We show that graphene, and other membrane-like monolayers, support a specific anharmonic connection between the flexural corrugation and longitudinal phonons whose fast escape leaves behind a 2D-projected mass density increase endowing the flexural phonons, as they move with their group velocity, with real momentum, part of which is transmitted to the adsorbate through scattering. The resulting distance-independent ballistic thermophoretic force is not unlikely to possess practical applications.
\end{abstract}

\maketitle

Thermophoresis is the phenomenon by which a body immersed in a fluid endowed with a temperature gradient experiences a force and, independent of convection, drifts from hot to cold \cite{duhr2006molecules}. We address here the less common case of thermophoresis of a physisorbed nano-object caused by an in-plane temperature imbalance
in the underlying solid substrate surface. Recent years have seen a surge of interest for methods to control nanoscale transport and manipulation, also in view of potential applications in nano-devices. The possibility to drive directional motion of adsorbates by means of thermal gradients is interesting and has been explored both theoretically \cite{schoen2006nanoparticle} and experimentally \cite{barreiro2008subnanometer}. By a similar principle the controlled directional motion on graphene was also explored by means of strain or wettability gradients \cite{wang2015motion, huang2014directional,liu2015actuating}. Carbon systems such as graphene and carbon nanotubes (CNTs) are prime candidate substrates \cite{barreiro2008subnanometer, schoen2007phonon, becton2014thermal, schoen2006nanoparticle, zambrano2008thermophoretic, becton2014thermal, shiomi2009water, savin2012transport} for these phoretic phenomena, owing to the their remarkable mechanical strength and thermal conductivity. Computational studies have highlighted the possibility to drive thermally gold nanoparticles, water clusters, graphene nanoflakes, C60 clusters, and small CNTs over graphene layers or inside CNTs. In spite of that, there appears to be so far insufficient intimate understanding of that phenomenon, besides the obvious consensus that the driving force stems from the spatial non-uniformity of the phonon population \cite{barreiro2008subnanometer, becton2014thermal, prasad2016phonon}. 
Our scope  will be to deepen that understanding.

To begin with, in a macroscopic system the heat current density $J$ is proportional to the temperature gradient (Fourier's law) through the thermal conductivity $\kappa$:

\begin{equation}\label{Fourier}
J = - \kappa \nabla T ~~.
\end{equation}

Existing discussions of adsorbate thermophoresis similarly assume that the adsorbate mass current will similarly be proportional to the temperature gradient

\begin{equation}\label{Fick}
J_a = - K \nabla T,
\end{equation}

where $K > 0$ now also depends on adsorbate, substrate and temperature - but not on its gradient. According to Eqs.~\ref{Fourier}-\ref{Fick}, systems with different substrate lengths $L$ but same gradient $\nabla T$ should yield the same heat current and the same adsorbate current -- that is, $\kappa$ and $K$ should not depend on $L$. These macroscopic expectations will of course be necessarily borne out for sufficiently large systems where transport is diffusive. 

However, for system sizes smaller than or comparable with the mean-free-paths (MFP) $\lambda_i$ of the heat carriers (phonons in our case), the heat transport may instead turn from diffusive to ballistic, where phonons with MFP $\lambda_i$ larger than $L$ carry the heat. In graphene, $\lambda_i$ can extend to hundreds of nm \cite{ghosh2008extremely} or even more \cite{lindsay2014phonon}, making Eq.~\ref{Fourier} invalid at the nanoscale. Ballistic heat transport along with ballistic-diffusive crossover have been discussed experimentally \cite{bae2013ballistic} and theoretically \cite{fugallo2014thermal, barbarino2015intrinsic}. Collective phonon excitations have also been proposed \cite {fugallo2014thermal} with MFPs possibly larger than those of single phonons, which might render the pure diffusive regime of Eq.~\ref{Fourier} only attainable with samples of size 0.1--1\,mm \cite{xu2014length}. To describe the small size regime one can phenomenologically introduce a non-local generalization of Eq.~\ref{Fourier} by replacing the constant $\kappa$ with $\kappa(x-x')$, with the resulting convolution leading to the product of Fourier-trasformed quantities \cite{allen2016non}. We note that  $\kappa(x-x')$ will however in our case depend on system size, a point to which we shall return later.

The question which we address here is what will happen to thermophoresis in the small size and distance regime, in particular whether Eq.~\ref{Fick} would still be valid or not in that case.
As we shall see, a ballistic regime emerges for thermophoresis too at short distances, where Eq.~\ref{Fick} breaks down, and a new understanding is necessary. This understanding will be mandatory for thermally induced transport of matter at the nanoscale.

As a specific test case we consider here the thermophoretic force felt by a gold cluster physisorbed on a graphene sheet of length $L$ suspended between two baths at temperatures $\Delta T$ apart. Graphene has an extremely large heat conductivity, amongst the highest of any known material, with measured values ranging from 2600 to 5300 $Wm^{-1}K^{-1}$ \cite{balandin2011thermal, barbarino2015intrinsic}. Thermal conductivity of suspended graphene is known to be dominated by acoustical lattice vibrations (even if the electron contribution to the total heat conductivity, estimated initially to be as low as 1\% \cite{ghosh2008extremely}, might be underestimated especially in doped and in short samples \cite{kim2016electronic}). The acoustical lattice vibrations of free graphene are in-plane transverse (TA), in-plane longitudinal (LA) and out-of-plane flexural (ZA). The contribution of the latter, much lower in frequency and therefore much more populated, has been shown to dominate in suspended graphene \cite{lindsay2010flexural, wang2010thermal, lindsay2014phonon}.

The thermophoretic motion of gold nanoclusters in carbon nanotubes (CNT) has been associated to collective motions of the carbon atoms of CNT by Schoen et al. \cite{schoen2007phonon}, and the subnanometer motion of cargos adsorbed on suspended nanotubes has been observed in a thermal gradient \cite{barreiro2008subnanometer}. Very recently it was shown via phonon wave packet molecular dynamics simulations in CNTs that thermophoretic motion is driven by the scattering of LA phonons between the external and internal CNTs \cite{prasad2016phonon}. In all theoretical studies so far the temperature gradient $\Delta T$ was nonetheless considered as the relevant quantity which control thermophoresis as in Eq.~\ref{Fick}.

Our present non-equilibrium molecular dynamics (NEMD) study between $T_{hot}$ = 475K and $T_{cold}$ = 325K shows that in the nanoscale regime below graphene sheet lengths $\sim$150\,nm the accurately measured thermophoretic force acting on the gold clusters is not at all proportional to the temperature gradient. Instead, the force is found to depend basically only on the absolute temperature difference $\Delta T$ between heat source and sink, independently of the sheet size $L$ between them. That unmistakeable evidence of ballistic thermophoresis is, we further establish, associated with the known ballistic heat transport, caused chiefly by ZA flexural vibrations. The phoretic force arises due to scattering of these phonons on the adsorbed cluster which is immersed in the phonon flow. Momentum is tangibly transfered from these phonons to the cluster. 
That is surprising at first, since harmonic phonons are not supposed carry real momentum, that can only be carried by a moving mass.

As it turns out, a specific anharmonic process of flexural phonons in a flexible but nearly inextensible 2D membrane concentrates extra 2D-projected mass when it is corrugated, as the ZA phonons do. These phonons then carry real momentum as they move with the ballistic ZA phonon phase velocity. Some of that momentum is picked up by the adsorbed object which as a result is phoretically pushed from hot to cold regions. The overall ballistic thermophoresis of clusters proposed here, yet to be verified experimentally, appears to bear some similarity with that which enables fast diffusion of water clusters on graphene ripples \cite{ma2016fast}.

Our paper is organized as follows. First we describe the model system, gold clusters on a graphene sheet, and the simulation approach. The approximate temperature profile of suspended graphene clamped between two unlike temperatures is discussed, and some of its features related to the phonon flux as described by McKelvey-Shockley-type theory \cite{maassen2015steady}. Next, a gold cluster is adsorbed on graphene and initially shown to be freely diffusing in thermal equilibrium. After that, a left-to-right temperature difference $\Delta T$\,=\,$T_{hot}-T_{cold}$ is turned on in the graphene sheet, and phoretic motion of the cluster is readily observed in the simulation. To measure accurately the phoretic force, the actual cluster motion is subsequently harnessed by a harmonic spring, acting as a dynamometer.
The phoretic force so obtained and its dependence on $L$ is examined and found independent of $L$ up to at least 150\,nm, indicating ballistic thermophoresis. 

In order to gauge the ballistic flux of ZA phonons which appears as the driving agent of thermophoresis, the frequency-selected energy transmission spectrum of monochromatic flexural waves is examined. The ability of a ZA phonon mode to carry physical momentum is shown to occur as a result of the mass-carrying mechanism also associated with an anharmonically entangled LA mode, a mechanism best understood by viewing graphene as a nearly inextensible membrane. Finally, the thermophoretic force resulting from scattering on the adsorbed cluster of this ``ZA+LA'' complex is demonstrated. The gradient-independent character of this nanoscale phoretic force invites a short final discussion.

\section{System and methods}

Because the physical results to be reached in this work are entangled with technical aspects of the simulation, we find it best to describe the system and methods first thing here, rather than letting the reader wonder about them until later chapters.

Graphene is described by a C--C Tersoff potential, reparametrized to better reproduce the experimental phonon spectrum \cite{lindsay2010optimized}. A gold cluster ($N$\,=\,459) with internal fcc structure and truncated-octahedral shape was described with Au--Au interactions of the embedded atom method (EAM) type, modified through a smooth cutoff \cite{johnson1988analytic}. The cluster 36-atom (111) facet is physisorbed on graphene.
Interaction between gold and graphene atoms is assumed to be of Lennard-Jones type with $\varepsilon$\,=\,22\,meV and $\sigma$\,=\,2.74\,\AA, as parametrized by Lewis et al. \cite{lewis2000diffusion}, a choice meant to reproduce the gold-graphite corrugation, rather than the adhesion energy.

\begin{figure*}[t!]
 \centering
 \includegraphics[width=\textwidth]{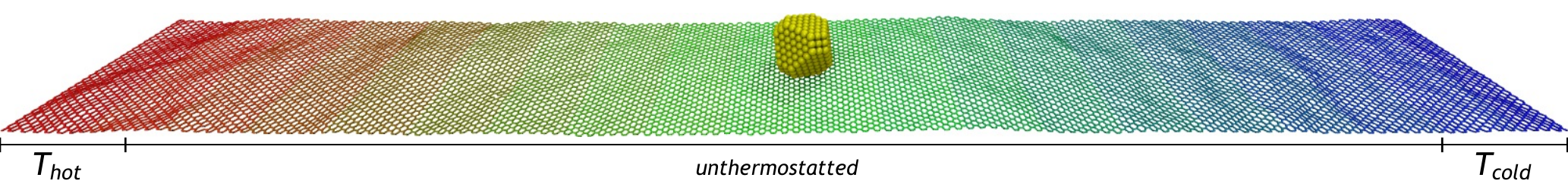}
 \caption{Schematic of an Au$_{459}$ cluster deposided on a 60$\times$7\,nm$^2$ graphene sheet,
 colored to highlight the thermal gradient. Thermostats are applied only on the $T_{hot}$ and
 $T_{cold}$ regions of graphene.}
 \label{fig:sketch}
\end{figure*}

We simulate a suspended graphene sheet of $x$-length $L$ and $y$-width $w \sim$\,6.5\,nm, with periodic boundary conditions along $y$. The first C-atom row ($x$\,=\,0) and the last one ($x$\,=\,$L$) are frozen, thus clamping the graphene sheet. A left-right temperature difference is introduced by coupling the first and the last 40 mobile atomic rows ($\Delta x\simeq$\,4.5\,nm) of the graphene sheet to two Langevin thermostats at temperatures $T_{hot}$ and $T_{cold}$, respectively (see Fig.~\ref{fig:sketch}). We typically use $T_{hot}$\,$\simeq$\,475\,K, $T_{cold}$\,$\simeq$\,325\,K, and a Langevin damping coefficient of 10\,ps$^{-1}$ strictly limited to the two thermostated regions, leaving the largest middle part of the graphene sheet (and the cluster when present) unthermostated. All initial equilibrium and subsequent NEMD simulations are conducted using our home-developed code. Two main approximations are the neglect of quantum effects, and of the finite-size acoustical phonon gaps at $k$\,=\,0. 

Quantum effects are absent in our entirely classical simulations. For that very reason (besides practical ones, including experimental accessibility) we choose to work at $T$\,$\sim$\,400\,K, a temperature that compromises between three constraints: 1) it is well above the temperature where the quantum effects of ZA flexural phonons (the dominant thermophoresis agent) become irrelevant \cite{bonini2012acoustic}; 2) it is $\sim$\,1/5 of the LA (and TA) Debye temperatures  \cite{pop2012thermal}, so that their specific quantum effect, although not irrelevant, are at least not dominant. LA and TA modes will anyway turn out not to contribute to thermophoresis; 3) it is still below the higher temperature regimes where some phonon mean free paths get anharmonically shorter than the system size \cite{chen2012thermal}.

The finite-size acoustical phonon gaps at $k$\,=\,0 on the other hand would be a problem if their magnitude came close to our working temperature $T$\,=\,400 K.  Even for the smallest size considered, $L$\,=\,30\,nm, however, the LA phonon gap $\sim$\,$v_{LA}$/2L is only $1.4$\,meV, or $16$\,K (for $v_{LA}$\,$\sim$\,22\,km/s); the ZA gap $\sim$\,C/4$L^2$ is for graphene $\sim$\,$5$\,$\times$\,$10^-3$\,meV or $60$\,mK. At $T$\,$\sim$\,$400$\,K both finite-size gaps are therefore irrelevant. 
 
\begin{figure}[tbh]
 \centering
 \includegraphics[width=0.4\textwidth]{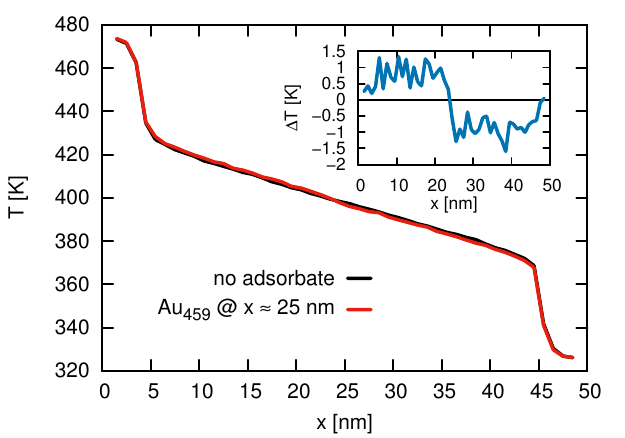}
 \caption{Local temperature profile in a graphene layer of length $L$\,=\,50\,nm, and width $w$\,=\,7\,nm, with and without the adsorbed gold cluster(located at $l$\,=\,22-24\,nm). The first (last) 4\,nm are thermostated at $T$\,=\,475\,K ($T$\,=\,325\,K). The temperature has been averaged over graphene slices of $1$\,nm along the gradient direction (see Method). In the inset the difference between the two temperature profiles is highlighted.}
 \label{fig:T_profile}
\end{figure}

The adsorbate-free graphene sheet is simulated first. For an approximate evaluation of the local temperature $T(x)$ we subdivide the graphene sheet in slices of 0.5\,nm along the gradient direction $x$, averaging the steady-state atomic temperature -- obtained by the equipartition theorem, $T$\,=\,$(2/3)N E_{kin}/k_B$ -- over the atoms in the slice and over a simulation time of at least 15\,ns, see Fig.~\ref{fig:T_profile}.

\begin{figure}[tbh]
 \centering
 \includegraphics[width=0.4\textwidth]{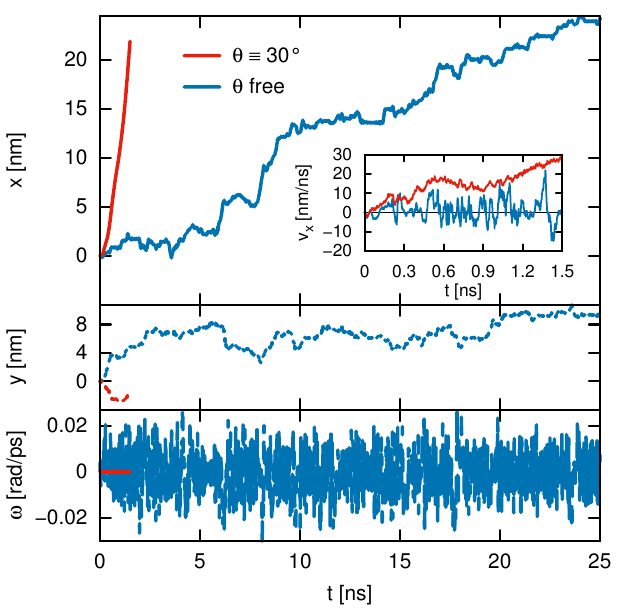}
 \caption{(top) The evolution of the position along $x$ of the center of mass of the Au cluster for a graphene length
 $L$\,=\,110\,nm with an applied $\Delta T$\,=\,100\,K between $T_{hot}$\,=\,450K and $T_{cold}$\,=\,350K.
 In the inset the velocity evolution for the first 1.5\,ns of the same simulations.
 (center) Cluster transverse displacemente along $y$ and (bottom) angular velocity $\omega$ during the same NEMD simulation.
 }\label{fig:free_ev}
\end{figure}

The cluster is then deposited near the center of the graphene sheet, following the adiabatic procedure described in Supporting Information \cite{SI}, a protocol which also permits the direct calculation of the (temperature-dependent) adsorbtion free energy in thermal equilibrium. In these equilibrium conditions, the cluster undergoes thermal diffusion, both positional and angular \cite{guerra2010ballistic}. Once the temperature difference $\Delta T $\,=\,$T_{hot}$\,-\,$T_{cold}$ is turned on, the cluster is observed to drift from hot to cold, as shown in Fig.~\ref{fig:free_ev}.

The approximate uniformity of the cluster center-of-mass (CM) motion indicates that the cluster-graphene friction is viscous \cite{guerra2010ballistic}. Given an average thermophoretic  force $F_{th}$ and assuming a viscous friction coefficient $\gamma$, the equation of motion for the average CM velocity v is

\begin{equation}\label{eq:Mdvdt}
M \frac{dv(t)}{dt} = F_{th} - \gamma v.
\end{equation}

Assuming the thermophoretic force to be sufficiently large to dominate fluctuations, the force can be extracted from the initial acceleration of the cluster from a rest condition with $v(0)$\,$\simeq$\,0 (see Supporting Information \cite{SI}).
Alternatively, once both the heat flow and the adsorbate drift reach the steady state regime, the mean velocity $\langle v\rangle(t \rightarrow \infty)$\,=\,$F_{th}/\gamma$ can be extracted \cite{barreiro2008subnanometer, schoen2006nanoparticle, schoen2007phonon, zambrano2008thermophoretic, rurali2010thermally}.

Neither method is very precise, however. Past work \cite{guerra2010ballistic} showed that $\gamma(T)$ is, in the diffusive regime, the time averaged result of a sequence of cluster jumps, rotations, and pinning events. The cluster needs a thermal fluctuation in order to misalign relative to the graphene lattice, and then it will positionally diffuse only when misaligned, events which occurs stochastically. Such erratic behavior, interesting as it is, requires very long simulation times, complicating the estimate of $\gamma$, and thus of $F_{th}$ in our case.
Some improvement can be obtained by artificially locking the cluster orientation $\theta$ relative to the underlying graphene lattice into a poorly commensurate state, e.g.\ at $\theta$\,=\,30$^\circ$, so as to obtain an artificially small but well-defined $\gamma$ and a large and smooth steady state speed suitable to an accurate extraction of $F_{th}$. The clear difference between the two phoretic regimes is visible in Fig.~\ref{fig:free_ev}.
Different $\theta$ values would lead to a different $\gamma$ and $\langle v\rangle$. Ignoring a possible angular dependence of momentum pickup rate by the cluster from the sheet, their product $F_{th}$ should be relatively independent of $\theta$ as desired.

\begin{figure}[tb]
 \centering
 \includegraphics[width=0.4\textwidth]{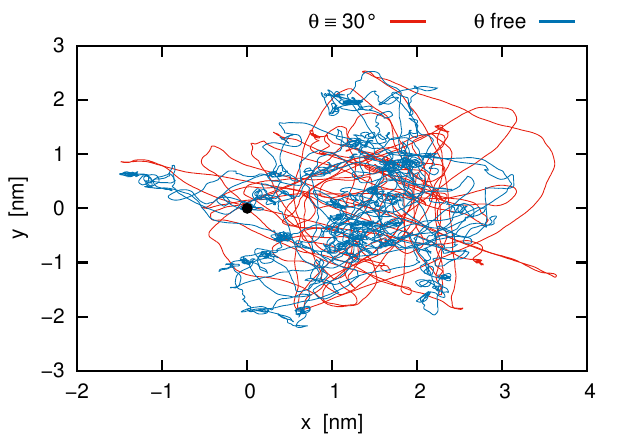}
 \caption{In plane trajectories for a spring-constrained gold cluster with (red) and without (blue) an imposed angular constraint subject to a temperature difference $\Delta T$\,=\,100\,K between $T_{hot}$\,=\,450K and $T_{cold}$\,=\,350K, on a graphene sheet of length $L$\,=\,50\,nm. The black spot corresponds to the rest position of the spring applied to the cluster center of mass. }
 \label{fig:trajs}
\end{figure}

In all cases however, the extraction of $F_{th}$ from the cluster trajectory is indirect, and thus tricky. It is therefore more convenient and immediate to measure directly the thermophoretic force $F_{th}$ from the NEMD simulation, by harnessing the mobile object, as also done by others \cite{becton2014thermal}.
We artificially tie a spring to the cluster CM by means of a potential term $(1/2)k_sX^2$, where $X$ is the cluster CM coordinate relative to the center of the sheet, and $k_s$\,$>$\,0 is a spring constant, strong enough to restrict the cluster motion to the most representative central (unthermostated) part of the system. Typical trajectories of the harnessed  cluster c.o.m. are shown in Fig \ref{fig:trajs}.
Experimentally, this kind of harnessed geometry might possibly be reproduced by an AFM tip, playing the role of our cluster, kept in place by a soft cantilever spring.

Owing to the modest cluster displacement permitted by the spring, $F_{th}$ can be obtained with the desired accuracy within a much shorter simulation time, $F_{th}$\,=\,$k_s{\langle}X{\rangle}$, where ${\langle}X{\rangle}$ is the average cluster CM displacement along the gradient direction. To further reduce the error and accelerate the averaging, the cluster alignment angle $\theta$ can be kept fixed by constraining $\theta$\,=\,30$^{\circ}$. 
Before systematically doing that, we checked that the angular constraint does not influence the calculated thermophoretic force (see SI for more info \cite{SI}). 

The statistical error affecting our results can be estimated both by the fluctuations of $F_{th}$ itself, $S_{X}$\,=\,$\langle k_s(X-\langle X \rangle)^2 \rangle$, and additionally checked by the fluctuations of the force along the direction perpendicular to the thermal gradient, $S_{Y}$\,=\,$k_s{\langle Y^2 \rangle}$. Since the position of the cluster has large autocorrelation times due to the slow dynamics, a block-analysis has been performed. 
Our overall error in this procedure is estimated to be $F_{err}$\,$\simeq$\,0.5--1.0\,pN, compared to  $F_{th}$\,$\simeq$\,2.0--10\,pN.
In order to enhance the resolution on the calculated average force we set the spring constant $k_s$\,=\,0.001\,meV/\AA$^2$\,$\simeq$\,16\,$\mu$N/m, a value
much smaller than the typical $\sim$1\,N/m of AFM cantilevers.
However, we note that for a stiffer spring the same resolution could be achieved by just increasing the simulation time.

\section{Temperature profile and heat flux}\label{sec.TprofileandHflux}

Let us begin with the results of a NEMD simulation of the free graphene sheet with the temperature imbalance, but without the adsorbed cluster.
Fig.~\ref{fig:T_profile} shows the typical average steady-state temperature profile. The thermal gradient that will become relevant later is the slope, obtained as a linear fit, in the central region. The steep temperature jumps near the two thermostated regions are a common feature found in all NEMD simulations \cite{zhou2009towards, cao2012kapitza,becton2014thermal}.
A reasonable explanation generally offered for the jumps is the same as the Kapitza resistance jump \cite{pollack1969kapitza}. The constrained left and right borders are mechanically different from the inside, and a travelling wave through that interface is partially reflected, causing a thermal resistance similar to that at the interface between two different materials \cite{ruraliPRB2014}.

In reality, the temperature jumps and their magnitude contain a much more interesting underlying source. Recent work by Maassen et al. \cite{maassen2015steady} has shown that near-border jumps in the effective temperature profile are present even for ``ideal'' contacts -- where no reflection occurs -- when at least some of the thermal current is ballistic, in the following sense.

In standard Boltzmann theory of thermal transport, the net heat current is the difference between forward and backward currents, $I_Q$\,=\,$I_Q^+$\,-\,$I_Q^-$. Following McKelvey \cite{mckelvey1961alternative} and Shockley \cite{shockley1962diffusion}, the scattering between forward and backward currents is controlled by a parameter, $\lambda$, roughly reflecting the phonon mean free paths. In this respect, the evolution of the heat currents inside the sample depends on the scattering probability between the two opposite currents. Considering only phonons with a given energy $\epsilon$,

\begin{equation}
\frac{\text{d}I^+_{Q}(x)}{\text{d}x} = \frac{\text{d}I^-_{Q}(x)}{\text{d}x} = -\frac{I_{Q}}{\lambda} ~~.
\end{equation}

The temperatures of the two thermostats enter as boundary condition in the values of the two currents. With an assumed ideal nature of the contacts, the right-flowing current through the left contact is in equilibrium with the left thermostat at $T_{hot}$, and similarly the left-flowing current through the right contact is in equilibrium at temperature $T_{cold}$, then

\begin{eqnarray}
I_{Q,0}^{+} = \epsilon \frac{M}{h} n(T_{hot}) \label{eq.coupled-fluxesI}\\
I_{Q,0}^{-} = \epsilon \frac{M}{h} n(T_{cold}) \label{eq.coupled-fluxesII}~~,
\end{eqnarray}

where $M(\epsilon)$ is the distribution of modes of the thermal conductor at energy $\epsilon$, $h$ is Planck's constant, $n(T)$ is the Bose-Einstein function. The coupled equations \ref{eq.coupled-fluxesI}-\ref{eq.coupled-fluxesII} for the currents can be solved to yield a value for the thermal gradient $\nabla T$ which is not, in general, equal to $\Delta T/L = (T_{hot}-T_{cold})/L$, thus implying the two temperature jumps at the two thermostated regions with ballistic transport variables.
The ballistic thermal resistance equals in fact $R^{ball} = I_Q / 2 \delta T$, where $\delta T$ is the temperature jump between thermostated-unthermostated regions, assumed to be the same at left or right contacts since the thermostat efficiency is fixed (given by the Langevin damping rate and by the extent of the thermostated area). Note that this resistance is present even for ideal, non-reflective contacts.

Summing up, the border temperature jumps represent \emph{the} prime evidence for  at least some ballistic heat transport. If all the heat current was transported ballistically, with zero dissipation from hot to cold, then both temperature jumps would be (in a symmetric case) $-\Delta T/2$, and ${\nabla}T$\,=\,0 in between. If on the contrary the heat flux was completely diffusive, then the jumps would disappear, Fourier's law would be obeyed throughout, and ${\nabla}T$\,=\,$-\Delta T/L$ everywhere. The temperature profile and the jumps in our simulated graphene are intermediate. The jumps are sizable, indicating a large amount of ballistic transport, 
besides some back-reflection.
The jumps are clearly smaller than $-\Delta T/2$, so that the average gradient ${\nabla}T$ is nonzero, corresponding to some amount of dissipation taking place in the middle region. As can be seen in the Supporting Information, the border jumps decrease consistently for increasing $L$, so that they will vanish for $L\rightarrow \infty$. 
With the new nanothermometric techniques ~\cite{menges2016temperature} it should be possible to verify this behavior.  

In view of the above, the nonlocality of heat conductivity $\kappa (x-x')$, very appropriately introduced by Allen ~\cite{allen} to account for the nonlinear temperature distribution in a thermal hetero-contact turns out in our case to be connected with the ballistic fraction of the heat flow: the range of $\kappa(x-x')$ varying from infinity to zero (i.e., locality) from the ballistic regime at small $L$ to the diffusive limit at very large $L$.

\subsection{Cluster adsorption}

Next, we adsorb on graphene the close-packed facet of a truncated octahedral $Au_{459}$ cluster. Prior to studying its thermophoretic motion, it is of interest to consider the adsorbtion thermodynamics and to characterize its thermal diffusion in full equilibrium, at zero temperature gradient. Following early experimental studies \cite{bardotti1996diffusion}, previous simulations of the diffusion of such gold clusters on graphene or graphite found a strong correlation between the rotational and translation degrees of freedom of the cluster \cite{ luedtke1999slip, lewis2000diffusion, guerra2010ballistic}.
At some angular orientations of the cluster the gold-graphene interface shows a strong interlocking, generally corresponding to minima of the zero-temperature adhesion energy (a more detailed discussion is reported in the SI \cite{SI}). At a majority of other orientations there is no interlocking, the adhesion is worse, and the cluster is much more mobile.
In this  "incommensurate" state the translational barrier drops dramatically, allowing positional diffusion to occur. The overall diffusive motion of the cluster consists therefore of an alternation of locked states at specific angles where only vibrations take place, and of diffusive states where the rotationally depinned cluster executes fast translations \cite{luedtke1999slip, guerra2010ballistic}.

A second feature of the adsorbed cluster to be understood before turning on thermal gradients is the cluster-graphene interface free energy $G(T)$, measuring the cluster adhesion as a function of temperature, in full thermal equilibrium. Unlike bulk free energies, the (negative) interface free energy may upon heating either rise (commonest) or drop (rare, but possible). In either case, should $G$ happen to depend strongly enough upon $T$, that dependence would in itself provide a source of thermophoretic force, once a spatial temperature gradient was introduced. We extract $G(T)$ from thermodynamic integration of simulation data, by the standard method of gradually removing the graphene-cluster interaction, passing from a fully interacting system to a purely free-standing cluster (more information in the SI \cite{SI}). The adsorption free energy $G(T)$, a negative quantity, is found here to weaken in magnitude with increasing temperature. Due to this dependence of $G$ on temperature, a thermodynamic force is calculated as $F_{ad}$\,=-\,$dG/dx$\,=\,-$dG/dT{\nabla}T \sim$\,0.29\,pN for ${\nabla}T$\,=\,1\,K/nm, a result that includes the small additional detachment of the cluster at higher temperatures.
As it turns out, this is 15 to 40 times smaller than the value necessary to account for the actual phoresis, to be described below. Even without any pretense to accuracy, this is far too small. Moreover, this contribution to the thermophoretic force scales by construction as ${\nabla}T$, which as we shall see is not compatible to our results. The origin of the main thermphoretic force must therefore be different.

\section{Cluster thermophoretic motion and force }

\begin{figure}[htbp]
 \centering
 \includegraphics[width=0.4\textwidth]{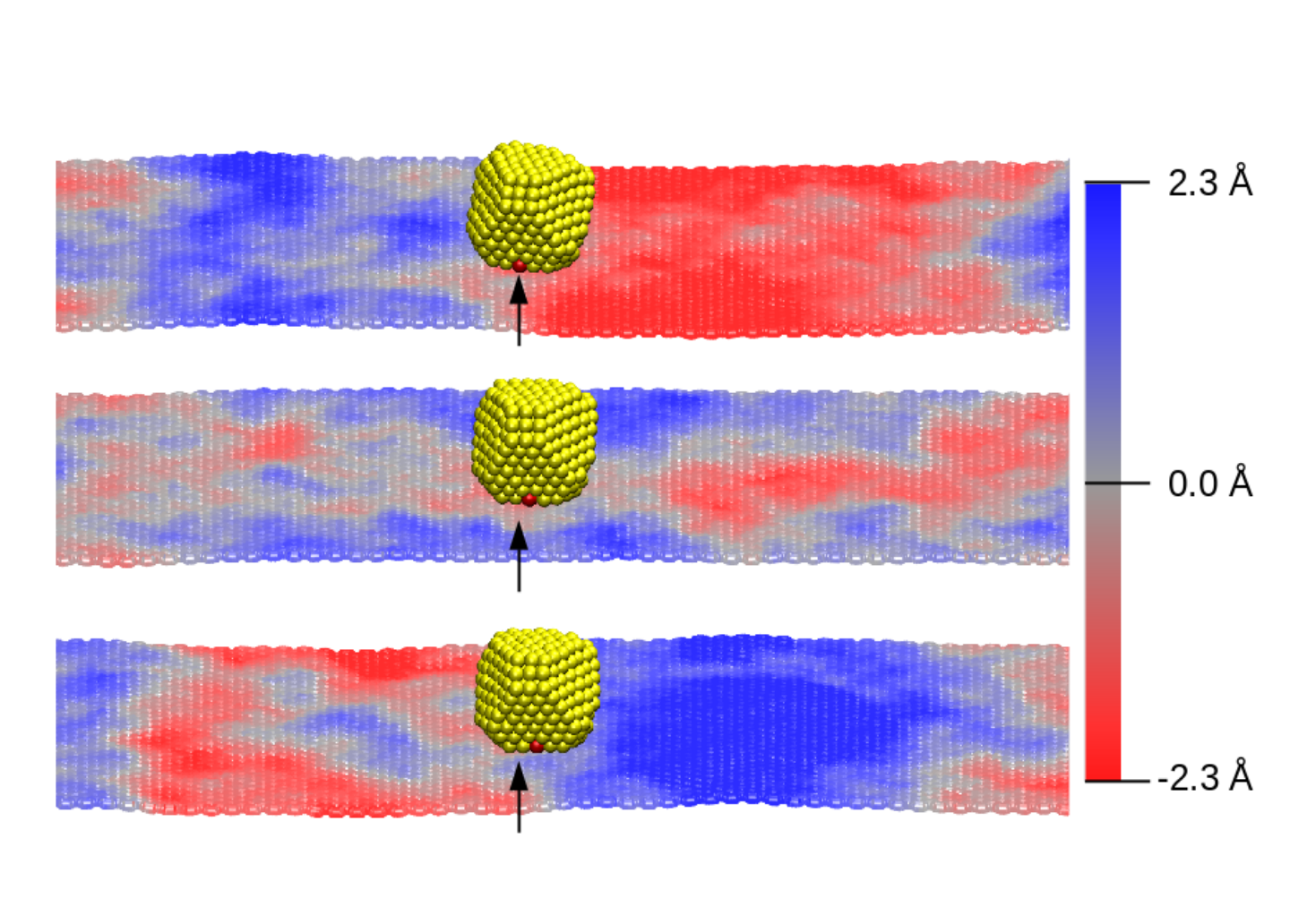}
 \caption{
A thermally excited flexural corrugation hits the cluster, pushing it along. These three snapshots (corresponding to a total $\Delta t$\,=\,15\,ps) have been obtained for pictorial purpose only, with a simulation where the left thermostat is set to $T_R$\,=\,700\,K and the right thermostat is set to $T_R$\,=\,0\,K so that only right-moving flexural phonons are visible. The black arrow indicates a fixed point in the graphene sheet, while one single atom of the gold cluster is colored in red to help keeping track of the forward motion of the cluster. The color code of the carbon atoms corresponds to their normal displacement relative to the perfect equilibrium plane.
}
 \label{fig:riding}
\end{figure}

By turning on the left-right temperature difference $\Delta T$ across graphene, we simulate thermophoresis. Firstly, as a purely visual and extreme illustration, we present in Fig.~\ref{fig:riding} a few snapshots where a thermal corrugation front, starting from $T_L$\,=\,700\,K, $T_R$\,=\,0\,K, hits the cluster from the hot region, carrying it along as if swept along by a "tidal" wave. 
Unrealistic as this sketchy situation is, it does provide an initial suggestion that  the ZA flexural vibrations are responsible for the thermophoretic motion.

More realistic simulations carried out by the standard protocol of Section~\ref{sec.TprofileandHflux} confirm the thermophoretic cluster drift, the X-distance as a function of time shown in Fig.~\ref{fig:free_ev}.
Based on that drift we estimate, using Eq.~\ref {eq:Mdvdt}, the thermophoretic force obtained from the evolution of the cluster velocity, with ${\Delta}T$\,=\,150\,K and with $\theta$\,=\,30$^{\circ}$ constrained so as to improve lubricity, as explained earlier. By fitting the initial acceleration of the cluster (see Supporting Information \cite{SI}) we obtain estimates of $F_{th}$ in the range  6.7$-$9.3\,pN. This direct approach is rather hard to converge for small sizes and not ideal for quantitative and accurate purposes, and as anticipated we proceed in the following to measure directly the average force by harnessing the cluster motion by a harmonic spring $k_s$. Measuring, as a in a dynamometer, the mean displacement ${\langle}X{\rangle}$ from the rest position at the midpoint of the sheet, the average thermophoretic force is extracted as $F_{th} = k_s{\langle}X{\rangle}$. After choosing a convenient $k_s$ value which roughly optimizes the time needed to determine the cluster displacement we carry out a systematic series of harnessed cluster simulations for increasing graphene sheet size $L$, two different temperature differences ${\Delta}T$, and two different graphene orientations, obtaining the spring-constrained thermophoretic forces reported in Table~\ref{tbl:ballistic}.

\renewcommand{\arraystretch}{1.0} 
\setlength{\tabcolsep}{8pt}       
\begin{table}[b!]
\begin{center}
 \begin{tabular}{cccc|c}
   $L$ (nm) & ${\nabla}T$ (K/nm) & ${\Delta}T$ (K) & ${\Delta}T^0$ (K) & $F_{th}$ (pN) \\
 \hline
   30       &  0.42              & 38              & 40                & 2.1 $\pm$ 0.5 \\
   40       &  0.39              & 37              & 40                & 1.7 $\pm$ 0.6 \\
   50       &  0.33              & 37              & 40                & 1.9 $\pm$ 0.7 \\
   60       &  0.29              & 37              & 40                & 3.6 $\pm$ 0.6 \\
   70       &  0.28              & 37              & 40                & 2.9 $\pm$ 0.8\\
 \hline
   30       &  1.88              & 148             & 150               & 9.7  $\pm$ 0.5\\
   40       &  1.47              & 147             & 150               & 9.2  $\pm$ 0.9\\
   50       &  1.26              & 147             & 150               & 8.9  $\pm$ 0.5\\
   60       &  1.10              & 147             & 150               & 10.4 $\pm$ 0.6\\
   70       &  1.04              & 147             & 150               & 10.1 $\pm$ 0.6\\
   150      &  0.54              & 147             & 150               &  8.8 $\pm$ 0.6\\
 \hline
   30       &  1.95$^\dagger$    & 147             & 150               &  9.0 $\pm$ 0.4\\
   40       &  1.71$^\dagger$    & 147             & 150               &  9.8 $\pm$ 0.6\\
   50       &  1.27$^\dagger$    & 147             & 150               &  9.3 $\pm$ 0.5\\
   60       &  1.15$^\dagger$    & 147             & 150               &  8.7 $\pm$ 0.4\\
   70       &  1.03$^\dagger$    & 147             & 150               &  7.9 $\pm$ 0.5\\
\end{tabular}
\caption{
  Average thermophoretic force $F_{th}$ evaluated for a $Au_{459}$ cluster deposited on a $L\times7$\,nm$^2$ graphene substrate with average temperature ${\langle}T{\rangle}$\,= $(T_{hot}+T_{cold})/2$\,=\,400\,K, and nominal temperature difference ${\Delta}T^0$\,=\,$T_{hot}$\,-\,$T_{cold}$ of 40\,K and 150\,K at the edges (the effective temperature difference is instead calculated as ${\Delta}T = T_{x=0}-T_{x=L}$). The local thermal gradient at the cluster site $\nabla T$ is also given (linear fitting of the central region, see Fig.~\ref{fig:T_profile}). The $\dagger$ symbol refers to thermal gradient applied in the zig-zag direction, while all the other simulations have thermal gradients applied in the armchair direction. }\label{tbl:ballistic}
\end{center}
\end{table}

The thermophoretic force $F_{th}$ displays, within tolerable errors, a linear dependence on the overall temperature difference ${\Delta}T$ but, remarkably, no dependence on the local gradient ${\nabla}T$ as would be expected from Eq.~\ref{Fick}, which is valid in the diffusive regime. This is a striking hallmark of ballistic behaviour. Even at our largest graphene sheet size $L$\,=\,150\,nm the expected decline of $F_{th}$ is still insignificant.

We conclude that up to at least the sizes considered in this work the thermophoretic force on absorbed nano-object on graphene remains largely ballistic.
In turn, this matches our previous observation that heat flux has a large ballistic component as indicated by the graphene temperature profile. The dependence of the force on ${\nabla}T$ generally invoked in literature even for nanoscale sizes is therefore not borne out here.

The close parallel with ballistic heat transport of graphene suggests that the ZA flexural vibrations are the actors responsible for the thermophoretic force. To check that, we repeat simulations by artificially freezing the out-of-plane graphene motion, only allowing in-plane LA and TA vibrations. In that case we obtain, for $L$\,=\,50\,nm and ${\Delta}T^0$\,=\,150\,K, a thermophoretic force of 0.5$\pm$0.7\,pN, negligible in comparison with the force found with unconstrained graphene, and more comparable to the error. 

We conclude that the flexural ZA vibrations of graphene are the main source of thermophoretic force on the adsorbed cluster. This conclusion now invites a more detailed analysis.

\subsection{Scattering of monochromatic flexural wave packets}

We need to qualify the role of phonons in the observed ballistic thermophoresis. The connection between the thermophoretic force and the atomic vibrations which transport the heat is direct at the nanoscale sheet size, where phonons, whose mean-free path is larger, remain sufficiently well defined. The temperature difference gives rise to a net imbalance of phonon population between the two sides, $n_{hot}$ and $n_{cold}$. In coaxial nanotubes, Prasad et al.\ \cite{prasad2016phonon} showed how a phonon wave packet traveling in the outer (longer) tube scatters at the edges of the inner (shorter) tube, exchanging energy and momentum. This ``push'' from a single wave is felt identically whether it comes from the hot reservoir or from the cold reservoir. However, since the phonons that constitute the wave packets from the two opposite regions have different populations a net phoretic force arises. In the ballistic transport regime, where Eq.~\ref{Fourier} is violated and $\delta n$ dominates, the latter is a function only of $\Delta T$, and so is the thermophoretic force resulting from this unbalanced population. In our case, the ballistic propagation of flexural phonons is identified as the source of the thermophoretic force. Harmonic phonons however should carry crystal momentum but no physical momentum: how can then the flexural phonons cause a net force?

We address this issue by simulating the time evolution of nearly monochromatic flexural phonon wave-packets. A phonon is injected in a free graphene sheet by applying an external vertical force on the left side of the non-thermostated region, shaking along $z$ a single column of carbon atoms with a force $F_{ex}$\,=\,$F_0\sin(\omega t)\hat{z}$ for a certain number of cycles $n_c$.
The shaking produced two symmetric flexural wave packets, a left-traveling one which enters the left thermostat and gets dissipated, and another right-travelling across the graphene sheet, finally absorbed by the right thermostat. By measuring the heat $W$ absorbed by the right thermostat, alternatively without and with the adsorbed cluster, we calculate the variation of transmitted energy $\Delta W(\omega)$ caused by the cluster. For each flexural frequency $\omega$ the heat transmission coefficient $\mathcal{T}$\,=\,$W_{cluster}/W_{clean}$ is extracted in this manner. To allow for a computationally viable and sufficiently precise calculation we study running waves of just a few ($n_c$\,=\,3--5) oscillations, so as to avoid interference with the backscattered wave, identifying the transmitted and reflected parts, finally absorbed by the opposite thermostats. Although ideally one could in the same way measure the flux of momentum besides energy, that is in practice substantially more difficult, and not really necessary. First, by injecting a small number of oscillations the resulting force on the cluster is often insufficient to overcome the static friction, and the cluster just oscillates around its average position. The magnitude of the oscillation is related to the momentum transfer between the wave packet and the cluster, but the net momentum transfer $P_{avg}$\,=\,$M\langle v_{CM}\rangle$ is zero, absorbed in this regime by the lattice underneath.
Secondly the flux of energy and its variation are well defined even in the absence of adsorbates, and can always be measured with good precision, unlike the total momentum flux, which as we will show in the following, has multiple components and cannot be measured with the same precision. We will therefore stick to the energy transmission, whose analysis is nonetheless quite informative.

\begin{figure}[tbh]
 \centering
 \includegraphics[width=0.4\textwidth]{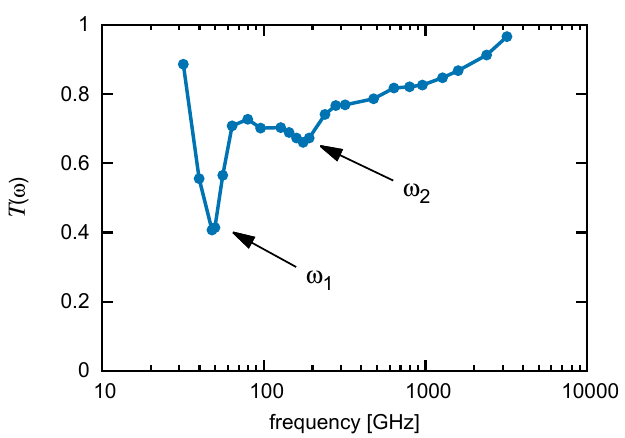}
 \caption{
   The flexural phonon energy transmission coefficient $\mathcal{T}(\omega)$ as a function of the phonon frequency. The low frequency resonances are discussed in the text.
 }
 \label{fig:transm}
\end{figure}

The phonon-resolved energy transmission of graphene sheets of tranverse size $w$\,=\,9\,nm and lengths in the range $L$\,=\,53--88\,nm is shown in Fig.~\ref{fig:transm}. At frequencies larger than $\sim$500\,GHz the transmittance increases smoothly from $\sim$0.80 to $\sim$0.97. The loss of transmission can be attributed to all processes that produce backscattering. 

In the low frequency region two transmittance dips are visible, approximately at $\omega_1$\,=\,49\,GHz and $\omega_2$\,=\,175\,GHz. These two features are connected with resonances between cluster-induced local vibrational modes and the incoming wave.
The first mode, $\omega_1$, coincides with the natural frequency of out-of-plane vibration of the cluster as a whole on the graphene substrate. This frequency depends on the Au--C interaction, and on the cluster size. Assuming approximately $\omega_1$\,$\sim$\,$\sqrt{K/M}$, a cluster mass $M$ and effective spring constant $K$ scaling proportional to $N$ and $N^{2/3}$, respectively, one obtains $\omega_1$\,$\propto$\,$N^{-1/6}$. Using the value obtained for $N$\,=\,459 we obtain $\omega_1$\,=\,$139 N^{-1/6}$\,GHz. The validity of this estimate is well verified for gold truncated octahedra clusters up to $N$\,=\,5635, where it yields $\omega_1'$\,=\,33\,GHz, close to the 31\,GHz value obtained in simulation.
This cluster $z$-mode $\omega_1$ has the same symmetry as the ZA flexural modes of graphene, and is therefore linearly coupled to that continuum.
The coupled problem actually provides an example of Fano resonance \cite{Fano1961}. The transmittance at a Fano resonance at $\omega$\,=\,$\omega_i$ is of the general form

\begin{equation}
T(\epsilon) = \frac{(\epsilon + q)^2}{(1+q^2)(1+\epsilon^2)}
\end{equation}

where $\epsilon$\,=\,$(\omega-\omega_i)/\Gamma$, $\Gamma$ being the width of the resonance and $q$ an asymmetry parameter, ranging from infinity (when $T(\epsilon)$ has a symmetric peak at $\epsilon$\,=\,0) to zero (when $T(\epsilon)$ has a symmetric dip at $\epsilon$\,=\,0).
The ZA phonon transmittance near $\omega_1$ can be fitted with $q$\,$\sim$\,0.0 and $\Gamma$\,$\sim$\,5 GHz.
To understand the second resonance, we recall that adsorption of the cluster on graphene is accompanied at $T$\,=\,0 by a local ``dimple'' of the graphene sheet under the cluster contacting face \cite{guerra2016slider}. We find that the dip frequency $\omega_2$ corresponds to a wavelength which is about twice the size of this dimple under the cluster. 
Thus, when the wavelength matches the size of the cluster contact facet the flexural phonon can more effectively set it into oscillation, and the incresed scattering maximum is reflected in a transmittance minimum. This kind of process may be connected to the recent observation that the fast diffusion of water drops on graphene is dominated by the graphene thermal ripples which have wavelengths large enough to ``contain'' the drops \cite{ma2016fast}.

At both dips in the transmission spectra, therefore, the resonant coupling between the incoming phonon and the cluster enhances the scattering process, contributing to the overall increase of the cluster-induced thermal resistance.
Interesting as it is, quantitative analysis however shows that the contribution of these resonances to the thermal resistance, and as a consequence to the thermophoretic force, is minor, compared to the remaining integrated continuum contribution over the whole frequency spectrum, which dominates.
By integrating the two on-resonance parts, we find that resonances only contribute 1.5\% of the total cluster-induced reduction of the transmitted energy.

In conclusion, all thermally excited flexural wavelengths, especially the longest, low-frequency  ones, contribute to the transmitted heat flux. The same occurs to the momentum transfer to the cluster, which we describe next. 

\subsection{Physical momentum of flexural traveling waves}

The energy-reflecting (and momentum-reflecting) nature of the cluster is first of all visible in the temperature profile shown in Fig.~\ref{fig:T_profile}: the cluster acts as a localized thermal resistance, resulting in a minute drop of 2\,K in the temperature profile.
The cluster acts as a scatterer for the incoming phonon packets, and picks up momentum that way.  

Here, as anticipated, we face a problem.
In a completely harmonic setting phonons carry crystal momentum but not physical momentum \cite{ashcroft1976solid}. At the harmonic level, there should therefore be zero net momentum exchange with the cluster even in presence of scattering, and zero ballistic thermophoretic force. This is not the case, and anharmonicity must be involved in the thermophoretic force.

Indeed, according to standard theory, a phonon $\omega_q$ is associated to a physical momentum $p_q$ (different from crystal momentum $q$) due to anharmonicity. 
If $\Gamma$\,=\,$-d\ln{\omega}/d\ln V$, is the anharmonic Gruneisen parameter, where $V$ is the volume, then $p_q$\,=\,$\Gamma\hbar q$ \cite{lewis1969note, physical-momentum-acoustic}. For the flexural phonons of graphene, the Gruneisen parameter is indeed large. Alas, it is \emph{negative} \cite{mounet2005first}, which at first is puzzling.  

While of course contained in the full derivations that can be found in the literature ~\cite{Michel2015anharmphon2D} the physical reason underlying the negative Gruneisen parameter of graphene, or any other membrane, is actually  simple and transparent. Thermally excited flexural phonons increase the total length of a clamped membrane causing a state of tension, therefore pulling the boundaries inwards - the contrary of normal bulk materials where temperature causes expansion. This connection between the negative Gruneisen parameter and the negative thermal expansion coefficient of graphene is well established \cite{schelling2003carbon}. 

If applied blindly to our case, standard theory would thus suggest that flexural phonons carry a negative physical momentum, at odds with the positive thermophoretic force observed in simulation. The solution of this puzzle is, as it turns out, quite interesting.
We find that a localized wave packet of flexural phonons in graphene conveys in reality a net \textit{positive} physical momentum, part of which is communicated to the adsorbate. This does not violate, as it would seem at first, the Gruneisen requirement because of the anharmonic association of the ZA phonon with a LA phonon of double frequency, longer wavelength and much larger group velocity. The \textit{combined} total momentum of the ZA-LA  phonon pair is negative as in Gruneisen theory. However the two wave packets separate and the flexural phonon, the only one that scatters the adsorbate, carries a positive physical momentum.

To establish that, we apply a vertical, time-dependent force $F_z = F_0 \sin(\omega t)$ to the row of graphene carbon atoms at position $X$\,$\simeq$\,20\,nm. This localized force creates two opposite-moving traveling flexural packets, whose oscillating parts are respectively of the form  $z^+$\,=\,$A\sin(q_{ZA}x - \omega t)$ and $z^-$\,=\,$A\sin(q_{ZA}x + \omega t)$ for $x>0$ and $x<0$. Consider now the forward-traveling phonon (the backward one behaves in exactly the same manner). 
Anharmonically associated with this harmonic ZA phonon there is an increase of the C--C bond lengths along $x$, since the ZA corrugation extends the effective length of clamped graphene by $\delta L/L \sim (q_{ZA}^2A^2/2) \cos^2(\omega t)$ (assuming $q_{ZA}A \ll 1$). 

This forced bond length modulation gives rise to a LA phonon wavepacket of double frequency $\tilde{\omega}$\,=\,$2\omega$, which travels away from $X$ with velocity $v_{LA}$ and a small wavevector $q_{LA}$ such that $ v_{LA} q_{LA} = 2\omega$. LA phonons are orders of magnitude faster than ZA phonons, and in extremely short times (t\,$\sim$\,1\,ps) the LA phonon carries away all longitudinal perturbations, leaving the remaining flexurally corrugated graphene with equilibrium C--C bond-lengths. The combination of corrugation and equilibrium bond-length produces a 2D projected density excess in the x-y plane $\delta\rho$, and therefore a positive physical momentum density associated the slow flexural phonon  $p$\,=\,$\delta \rho v_{ZA}$.

\begin{figure}[tbh]
 \centering
 \includegraphics[width=0.4\textwidth]{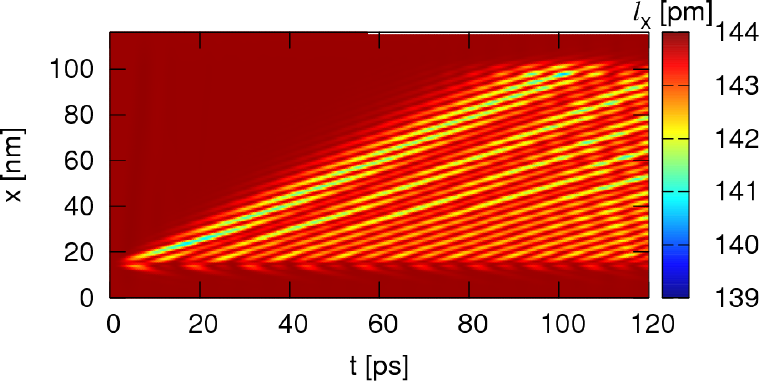}
 \caption{
 The  $x$-projected C--C  bond length  $l_x$, averaged over the $y$ direction,  mapped as a function of longitudinal position  $x$ and time, as the external force shakes the clean graphene sheet (no cluster)  at $X$\,$\simeq$\,20\,nm  with 80\,GHz frequency.  The excitation and propagation of  ZA flexural vibrations is visible. Note that there is basically no reflection when the ZA phonon hits the boundary where the thermostat damping begins. 
 }
 \label{fig:proj_len}
\end{figure}

\begin{figure}[t!]
 \centering
 \includegraphics[width=0.4\textwidth]{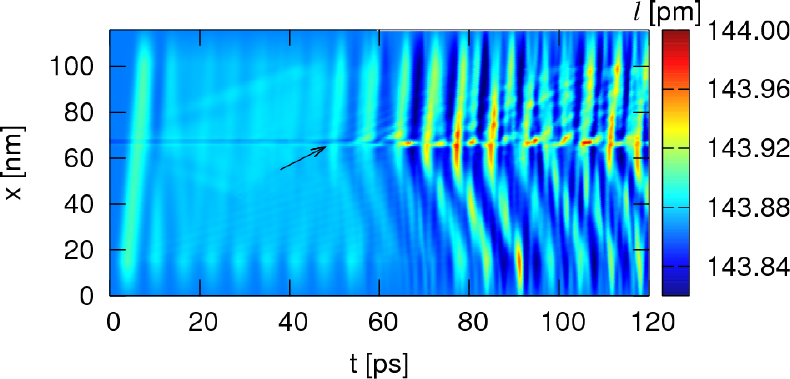}
 \caption{
 The total C--C bond length $l$, averaged over the the $y$ direction, mapped as a function of the longitudinal $x$ position and of time, as the external force shakes the graphene sheet $X$\,$\simeq$\,20\,nm with a physisorbed cluster located at $X$\,=\,65\,nm. The arrow indicates the point in time and space where the first flexural corrugation reaches the adsorbed cluster. Note how up to that time the presence of the cluster has little effect on the bond lengths.}
 \label{fig:true_len_cluster}
\end{figure}

Figs.~\ref{fig:proj_len} and \ref{fig:true_len_cluster} show as an example the case of $\omega$\,=\,80\,GHz and a z-shaking force $F_0$\,=\,30\,pN applied at $X$\,$\simeq$\,20\,nm of the suspended graphene sheet (no adsorbed cluster for now). Upon increasing the intensity $F_0$  it is verified that the ZA corrugation $A$ scales linearly with  $F_0$, accompanied  by an LA excitation amplitude which scales like $A^2$, as expected for its anharmonic origin. In the first picoseconds after switching on of the force we monitor the projected bond length, defined as the distance between carbon atoms projected on the x-y plane, and the true C-C bond-length. 

\begin{figure}[tbh]
 \centering
 \includegraphics[width=0.4\textwidth]{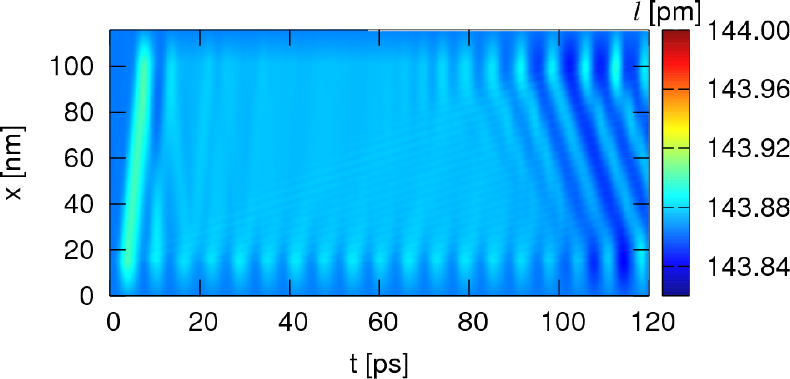}
 \caption{
The total  C--C bond length $l$, averaged over the the $y$ direction, mapped as a function of position  $x$ and time, as the external force shakes the clean graphene sheet (no cluster) $X$\,$\simeq$\,20\,nm with 80\,GHz frequency. The anharmonic excitation and propagation of the fast LA longitudinal mode is visible, accompanying the flexural mode of the preceding figure. Note that unlike the ZA mode, this LA mode undergoes visible reflection due to thermostat damping onset at the boundary. }
 \label{fig:true_len}
\end{figure}

In Fig.~\ref{fig:proj_len} the in-plane projected  $C$--$C$ bond lengths are reported as a function of position and time. The propagating flexural ZA phonon involves a projected bond length which is shorter than the static equilibrium value, indicating that graphene is slightly ``over-dense'' under the slowly moving mode of velocity $v$\,$\sim$\,0.9\,km/s , close to the theoretical value $d\omega_{ZA}/dq_{ZA}$\,=\,1.06\,km/s ~\cite{pop2012thermal}. The reason for that higher density is clarified in Fig.~\ref{fig:true_len}, where the true bond lengths are shown. At very short times there are  compensating ``under-dense'' modulations (bonds longer than the equilibrium value) which move away from the excitation point. Their apparent speed is $v$\,$\sim$\,22\,km/s, close to the experimental speed of sound for LA phonons $v_{LA}$\,$\sim$\,21\,km/s \cite{lindsay2010optimized}. Note the scale difference of the two pictures: while the true bond length modulation is of 0.03\%, the projected length shrinks by as much as 2\%; and this for all ZA frequencies $\omega_z$. The explanation for this mismatch lies in the fact that the modulations in the ZA and LA phonons are ``diluted'' over the relative wavelengths, which are very different: indeed $\lambda_{LA}\approx 138 nm$ and $\lambda_{ZA}\approx$\,7.9\,nm. To better compare the two contribution we can estimate the momentum of the two modulations produced in a cycle of the vertical force. The ratio, for the simulation discussed above, can be calculated as $P_{LA}/P_{ZA}\sim(2 \lambda_{LA}v_{LA}\delta\rho_{LA})/(\lambda_{ZA}v_{ZA}\delta\rho_{ZA})$\,$\sim$\,$-12$. This result indicates that the total momentum is indeed negative as Gruneisen's theory indicates (the factor 2 in the formula above comes from the double frequency).

Simulations thus confirm that the longitudinal modulation, although anharmonically created together with the flexural mode, moves away and separates very quickly. The remaining flexurally corrugated graphene has an increased mass density per unit projected area and its associated physical momentum is  positive. This mechanism bears a resemblance with the picture by Bassett \textit{et al.} \cite{bassett1966momentum} who investigated the physical momentum of localized running waves for 1D systems; to the best of our knowledge no further investigation of this mechanism was done following that work.

\begin{figure}[t!]
 \centering
 \includegraphics[width=0.4\textwidth]{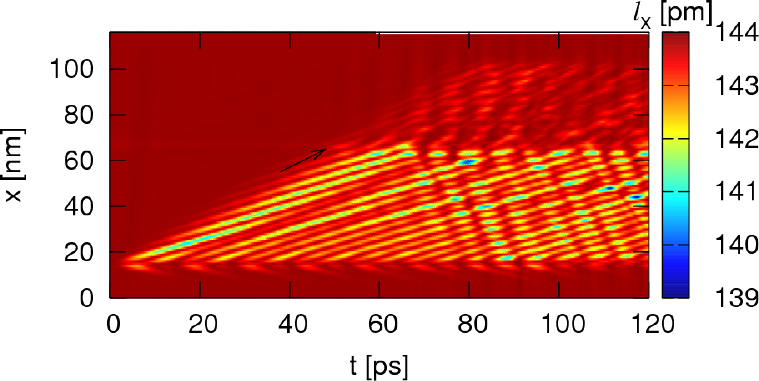}
 \caption{
 $x$-projected C--C graphene bond length $l_x$ averaged over the $y$ direction, mapped as a function of the longitudinal $x$ position and of time for a graphene sheet with a physisorbed cluster located at $X$\,=\,65\,nm, excited by a $z$ oscillation applied at $X$\,=\,20\,nm. The arrow indicates the point in time and space where the first flexural corrugation reaches the adsorbed cluster.
 }
 \label{fig:proj_len_cluster}
\end{figure}

Consider now the interaction of these artificially excited phonons with the gold cluster adsorbed in the middle of the graphene sheet. The cluster produces intense scattering of the flexural wave packets, as visible from the projected bond lengths (Fig.~\ref{fig:proj_len_cluster}). In this way, it picks up positive physical momentum, which explains the thermophoretic force. At the same time, there is no visible cluster-related  scattering of the fast-moving longitudinal phonons. The LA cross section on the cluster is very small as can be seen comparing the evolution of the true bond length in simulations with and without the adsorbed cluster, Fig.~\ref{fig:true_len_cluster} and Fig.~\ref{fig:true_len}: indeed the presence of the cluster has little effect until the ZA wave packet hits the cluster and additional anharmonic effects take place. As a result, the  negative LA momentum is not transferred to the cluster.

\subsection{Phoretic force by a flexural phonon}

We are now in a position to present the resulting formulation for the phoretic force caused by a single injected flexural phonon in graphene, valid also for a more general membrane-like 2D material.
While (predominantly) propagating from hot to cold, each flexural phonon en route impinges on the adsorbed cluster, which picks up momentum by parly reflecting it backwards. This can be described as follows.

The excess density embedded in a traveling ZA wave moves with its phase velocity. This excess projected density $\delta \rho_C$ is 

\begin{equation}
\delta \rho_C / \rho_C = (l_{eq}/l_x - 1)
\end{equation}

where $l_x$ is the projection along $x$ of the average carbon-carbon bonds length whose equilibrium length is $l_{eq}$. The physical momentum density $p$ per unit area is therefore:

\begin{equation}
p_q = v_q \rho_C (l_{eq}/l_x - 1)
\end{equation}

It should be stressed again that this physical $x$-momentum is not directly produced by the external force which generates the flexural phonon, a force that acts only along $z$. The momentum is an anharmonic effect, proportional to the square of the ZA phonon amplitude $A$, and the result of the anharmonic pairing of the ZA and an LA phonon of very different wavevector and velocity. 

The phoretic force $F_{th} = \frac{dP}{dt}$ produced by partial back-reflection by the the adsorbate of contact y-width $h$ of a wavepacket of wavevector $q$, phase velocity $v_q$ , amplitude $A$, and transmission coefficient $1-\alpha_q$, is

\begin{equation}
F_{th} \simeq  2 \alpha_q (h v_q) p_q \\
       = 2 \alpha_q h \rho_C (l_{eq}/l_x - 1)  v_q^2
\end{equation}

where the factor 2 accounts for backwards reflection of the scattered phonon (1D behavior assumed for simplicity). 
 
To check this result in a specific case we carry out a specific simulation of the suspended graphene sheet of $L$\,=\,85\,nm with the adsorbed cluster at $X$\,$\simeq$\,50\,nm by applying a z-shaking force $F_0$\,=\,50\,pN applied at $X$\,$\simeq$\,20\,nm and  frequency $\omega$\,=\,80\,GHz. Here we can compare the above prediction with the phoretic force as measured by the initial acceleration of the free cluster.

From an independent, cluster-free simulation with the same z-shaking force $F_0$\,=\,50\,pN we extract - after reaching  steady state - an average $\delta\rho_C/\rho_C$\,=\,1.5\%. Assuming the momentum transmission coefficient $1-\alpha_q$ to be well approximated by the energy transmission coefficient extracted earlier on, $\alpha_q$\,=\,0.28\,=\,$1-\mathcal{T}(\omega=80\,GHz)$, we obtain a phoretic force of $F_{th}$\,$\sim$\,20\,pN, in reasonably good agreement with the actual force acting on the cluster of 25\,$\pm$\,2\,pN, directly obtained in simulation.
In principle, by repeating the procedure for all ZA modes and integrating over the thermal distribution with its left-right imbalance one will approximate the total thermophoretic force. That however would demand a massive effort which we do not undertake. Since the single phonon force calculation works, there is no reason to doubt that the overall integrated thermophoretic  force would also work.

\section{Conclusions}

An externally imposed temperature difference $\Delta T$ between the extremes of a vacuum suspended graphene sheet is predicted to push thermophoretically an adsorbed cluster from hot to cold. For submicron nanoscale sheet sizes the theoretical thermophoretic force is proportional to $\Delta T$ but independent of sheet length $L$ and thus independent of the thermal gradient $\nabla T$, a key evidence of ballistic phoresis. The main agents of this effect are the flexural phonons, whose long mean-free path and ability to transport heat ballistically are well known. Besides heat, flexural phonons also carry physical momentum, owing to their membrane-like increase of projected density. In turn, that is associated with the special anharmonic coupling to a longitudinal phonon that takes the compensating density decrease away from the scattering region. The flexural phonon flux flowing from hot to cold cedes some of its physical momentum by scattering onto the adsorbed cluster, which is ballistically trasported.
Previously known examples of ballistic phoresis include the Knudsen force exerted by gas particles on tips, described by Passian \cite{passian2002knudsen} and observed by Gotsmann \cite{gotsmann2005experimental}.
The phonon-induced  ballistic thermophoresis described here is specific to submicron sheet sizes and will eventually disappear as flexural phonons evolve from ballistic to diffusive once the sheet length $L$ is large enough, a crossover which our simulations do not yet detect at $L \sim 150 nm$. Below that crossover size, the possibility to realize a distance-independent ballistic force is remarkable, and not unlikely to find practical applications. 
As an example the ballistic character of thermophoresis on graphene could allow long-range, non-contact action by a moving heat source such as a hot cantilever \cite{menges2012quantitative} on adsorbed clusters or molecules. Other current uses of the thermophoretic effect at micro- and nano-scale, i.e. particle separation \cite{lervik2014sorting} or evaluation of molecular interactions \cite{jerabek2011molecular}, could also benefit from the ballistic regime.
Sensitive tools such as so-called ``pendulum'' Atomic Force Microscopes \cite{kisiel2011,langer2014} or Quartz Crystal Microbalances \cite{krim1988damping, bruschi2001measurement} should be able to detect this effect, and to characterize the expected ballistic-diffusive crossover.

\section*{Acknowledgments} Study conducted under ERC Advanced Grant 320796 MODPHYSFRICT, also partly sponsored by European COST Action MP1303.

\bibliography{biblio_thermo}

\end{document}